\documentclass[aps,pre,reprint,superscriptaddress,amsmath,showpacs,amssymb]{revtex4-2}

\usepackage[T1]{fontenc}

\usepackage{epsfig}
\usepackage{graphicx}
\usepackage{bm}
\usepackage{color}
\usepackage[hyperindex,breaklinks,unicode,colorlinks=true,linkcolor=blue,citecolor=blue,urlcolor=blue]{hyperref}

\usepackage[draft]{changes}

\newcommand{\slow}{{\rm slow}}

\newcommand{\be}{\begin{equation}}
\newcommand{\ee}{\end{equation}}

\definecolor{myblue}{rgb}{0.2,0.2,0.8}
\definecolor{myblack}{rgb}{0,0,0}

\definecolor{zhuolinblue}{rgb}{0,0.5,0}

\begin{document}


\newcommand\uleip{\affiliation{Institut f\"ur Theoretische Physik, Universit\"at Leipzig,  Postfach 100 920, D-04009 Leipzig, Germany}}
\newcommand\chau{\affiliation{ 
 Department of Macromolecular Physics,  
 Faculty of Mathematics and Physics, Charles University,
 V Hole{\v s}ovi{\v c}k{\' a}ch 2, 
 CZ-180~00~Praha, Czech Republic
}}
\newcommand\icfo{\affiliation{ 
ICFO—Institut de Ci{\` e}ncies Fot{\` o}niques, The Barcelona Institute of Science and Technology, 08860 Castelldefels (Barcelona), Spain
}}
\newcommand\geneva{\affiliation{ 
Department of Applied Physics, University of Geneva, 1211 Geneva, Switzerland
}}
\newcommand\exeter{\affiliation{Department of Physics and Astronomy, University of Exeter, Stocker Road, Exeter EX4 4QL, UK
}}
\newcommand\oxford{\affiliation{ 
Department of Materials, University of Oxford, Parks Road, Oxford OX1 3PH, United Kingdom
}}
\newcommand\potsdam{\affiliation{Institut f\"ur Physik und Astronomie, University of Potsdam, 14476 Potsdam, Germany
}}

\title{Optimal finite-time heat engines under constrained control}

\author{Zhuolin Ye}\email{zhuolinye@foxmail.com}\uleip

\author{Federico Cerisola}\oxford\exeter

\author{Paolo Abiuso}\icfo\geneva

\author{Janet Anders}\exeter\potsdam

\author{Mart{\' i} Perarnau-Llobet}\geneva

\author{Viktor Holubec}\email{viktor.holubec@mff.cuni.cz}\chau


\begin{abstract}
\textcolor{black}{We optimize finite-time stochastic heat engines with a periodically scaled Hamiltonian under experimentally motivated constraints on the bath temperature $T$ and the scaling parameter $\lambda$. We present a general geometric proof that maximum-efficiency protocols for $T$ and $\lambda$ are piecewise constant, alternating between the maximum and minimum allowed values. When $\lambda$ is restricted to a small range and the system is close to equilibrium at the ends of the isotherms, a similar argument shows that this protocol also maximizes  output power. These results are valid for arbitrary dynamics. We illustrate them for an overdamped Brownian heat engine, which can experimentally be realized using optical tweezers with stiffness $\lambda$.}
\end{abstract}


\maketitle


\section{Introduction}  The unprecedented improvement in experimental control
over microscopic Brownian~\cite{Pesce2020} and quantum
systems~\cite{Wineland2013,Haroche2013,myers2022quantum} has induced a revolution in the study of
 heat engines~\cite{Sekimoto2010,seifert2012stochastic}. It aims to generalize equilibrium and finite-time thermodynamics~\cite{curzon1975,Rubin1979,Salamon1980,Andresen1984,Bjarne1984,Mozurkewich1986,Broeck2005,Cisneros2007,Hoffmann2008} to the nanoscale, where thermal and quantum fluctuations render
thermodynamic variables such as work and heat
stochastic~\cite{holubec2021fluctuations}. Intense effort is devoted
to uncover optimal performance of stochastic {heat} engines~\cite{Geva1992,Parmeggiani1999,Hondou2000,Feldmann2000,Astumian2002,Parrondo2002,Reimann2002,Schmiedl2008,Tu2008,Esposito2009,Esposito20091,Esposito2010,Abah2012,Raz2016,Holubec2018,PhysRevLett.124.040602,holubec2021fluctuations,alonso2021geometric,cavina2021maximum,erdman2019maximum,PhysRevResearch.2.033083,esposito-EMP-bounds,Abiuso2020,PhysRevLett.89.116801,abiuso2020geometric,PhysRevLett.126.210603,slowdriving}. 
However, optimal control protocols 
are only known {under  approximations of
fast~\cite{cavina2021maximum,erdman2019maximum,PhysRevResearch.2.033083} or
slow~\cite{esposito-EMP-bounds,Abiuso2020,PhysRevLett.89.116801,abiuso2020geometric,PhysRevLett.126.210603,slowdriving}
driving, or for specific microscopic models:} 
engines based on overdamped Brownian particles in
harmonic~\cite{Schmiedl2008}, log-harmonic~\cite{Holubec2014}, or
slowly-varying potentials~\cite{SI}, and for an underdamped harmonic Brownian
heat engine~\cite{Dechant2017}. {Furthermore, most these} 
{exact results} have been obtained under constraints on the state of the working medium of the engine~\cite{Zhang2020}, instead of experimentally motivated constraints on the control parameters~\cite{Abiuso_2022,Zhong2022}. \textcolor{black}{An exception is Ref.~\cite{PhysRevE.93.042112}, showing that reaching maximum efficiency of slowly driven cyclic heat engines requires control over the scaling of the full Hamiltonian to avoid heat leakages.} 

\textcolor{black}{
In this letter, we optimize finite-time
thermodynamic cycles under constraints on control parameters such as trap stiffness of optical tweezers $\lambda$ and bath temperature $T$. We show that, different from constraining the response such as the width $\sigma$ of the phase distribution, constraining the control allows for surprisingly simple and general derivation of maximum-efficiency and maximum-power protocols. Besides other stark differences, for constrained control of Brownian heat engines, these protocols significantly outperform the protocol optimised for power and efficiency under constraints on $\sigma$~\cite{Schmiedl2008}. 
}

\section{Setup} 
\textcolor{black}{
Following Ref.~\cite{PhysRevE.93.042112}, we} assume that the Hamiltonian of the system that serves as a working
medium of the stochastic heat engine is of the form,
\begin{equation}
  H(x,t)= \lambda(t) f(x),
  \label{eq:general-halm}
\end{equation}
where the control parameter $\lambda(t)$ periodically expands and shrinks the
energy spectrum in time, and \textcolor{black}{$f(x)$ is an arbitrary function of the system degrees of freedom $x$ such that the equilibrium partition function $Z(t)=\int dx \exp[-H(x,t)/(k_{\rm B}T)]$ is finite for all $k_{\rm B}T \ge 0$.}  \textcolor{black}{This class of Hamiltonians 
generalizes the well-known `breathing'
parabola model~\cite{Schmiedl2008} for an overdamped particle trapped in a parametrically driven harmonic potential. It also includes semi-classical two (or multi) level systems with controlled gaps between the individual energy levels~\cite{holubec2021fluctuations}, and quantum spins, where the control parameter is an externally controlled magnetic
field~\cite{Geva1992}.}

We connect the system to a heat bath and periodically alter its
temperature $T(t)$ with the same finite period $t_{\rm p}$ as $\lambda(t)$. The
parameters under experimental control are thus $\lambda(t)$ and $T(t)$, and we
assume that experimental conditions restrict their minimum and maximum values \cite{PhysRevE.99.012140}.  
Our aim is to find optimal $t_{\rm p}$-periodic protocols for the constrained
control \textcolor{black}{parameters 
\begin{equation}
\label{eq:constraints}
\lambda(t) \in [\lambda_-, \lambda_+],\quad
T(t) \in [T_-, T_+].
\end{equation}
}

\begin{figure}[t]
\centering
\includegraphics[trim={0cm 0.2cm 0cm 0cm},width=0.67\columnwidth]{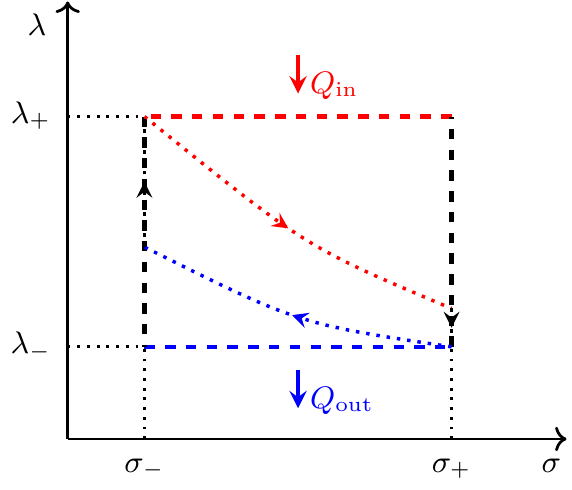}
\caption{\textcolor{black}{The maximum-efficiency protocol \eqref{eq:maxEffprot} under the constraints in Eq.~\eqref{eq:constraints} (dashed line) compared to a suboptimal cycle (dotted line). Heat flows into the system during the red branches of the cycles ($d\sigma>0$) and is released to a heat sink during the blue branches ($d\sigma<0$).}}
   \label{fig:setting}
 \end{figure}

\section{Maximum efficiency}
\textcolor{black}{Our first main result is a general geometric proof that the maximum-efficiency finite-time cycle under the constraints~\eqref{eq:constraints} is a Carnot-Otto cycle composed of two isotherms/isochores interconnected by two adiabats. The maximum-efficiency protocol $\{T(t),\lambda(t)\}$  is thus piecewise constant:
\begin{equation}
\label{eq:maxEffprot}
\{T(t),\lambda(t)\}_{\eta} = 
    \begin{cases} 
      \{T_+,\lambda_+\}, & 0 < t < t_{+} \\
      \{T_-,\lambda_-\}, & t_{+} < t < t_{\rm p}
   \end{cases}.
\end{equation}
And the maximum efficiency is given by
\begin{equation} \label{eq:opteff}
  \eta = 1 - \lambda_{-}/\lambda_{+}.
\end{equation}
The proof relies just on the definition of heat and it is thus independent of the details of the system dynamics, including the times $t_{+}$ and $t_{\rm p}$. It holds both for situations when the heat bath is memoryless (Markovian) and non-Markovian. The non-equilibrium dynamics of the system communicating with a Markovian bath can be described by Fokker-Planck or master equations for the probability density for $x$~\cite{Risken1996}. Except for a few exactly solvable settings~\cite{holubec2021fluctuations,Risken1996}, these equations are usually hard to solve analytically for non-quasi-static time-dependent protocols. However, in the non-Markovian case, a corresponding closed deterministic description might not be available at all~\cite{Holubec2022}. Then one has to resort to stochastic descriptions, such as a generalized Langevin equation, making even a numerical optimization challenging. The derivation also holds in situations with a non-equilibrium bath, such as in recently intensely studied cyclic active Brownian heat engines~\cite{Krishnamurthy2016,HolubecActive,Kumari2020,Gronchi2021}.}

\textcolor{black}{
Let us now derive Eqs.~\eqref{eq:maxEffprot}-\eqref{eq:opteff}.}
Under reasonable assumptions, any periodic variation of the control
parameters eventually induces a periodic average response of the system,
$\sigma(t) = \langle f[x(t)] \rangle$. This \textcolor{black}{ensemble average} is a functional of $T(t)$ and $\lambda(t)$ specified by dynamical equations of the system.
\textcolor{black}{Due to the factorized structure of the Hamiltonian \eqref{eq:general-halm}}, the average internal energy of the system $\langle H(x,t) \rangle$ is given by
$\lambda(t)\sigma(t)$. Decomposing its infinitesimal change into a component
corresponding to the external variation of the control $\lambda$ (work) and the
rest (heat)~\cite{seifert2012stochastic,Sekimoto2010}, it follows that output
work and input heat increments are given by \dj $W_{\rm out}(t) = - \sigma(t)
d\lambda(t)$ and \dj $Q(t) = \lambda(t) d\sigma (t)$, respectively. Per cycle,
the engine transforms the fraction $\eta = W_{\rm out}/Q_{\mathrm{in}} = 1 -
Q_{\mathrm{out}}/Q_{\mathrm{in}}$ of the heat $Q_{\rm in} = \int_0^{t_{\rm p}} \lambda(t)
\theta[d\sigma (t)] d\sigma (t)$ from the heat source into output work $W_{\rm
out} = -\int_0^{t_{\rm p}} \sigma(t) d\lambda(t)$, and dumps the remaining heat $Q_{\rm
out}=Q_{\rm in}-W_{\rm out} = \int_0^{t_{\rm p}} \lambda(t) \theta[-d\sigma
(t)] d\sigma (t)$ into the heat sink ($\theta(\bullet)$ denotes the
Heaviside step function \textcolor{black}{reflecting that heat flows on average into the system when $\sigma$ increases, i.e., $d\sigma>0$}).

Consider now the $\lambda-\sigma$ diagram of the cycle depicted in
Fig.~\ref{fig:setting}. \textcolor{black}{We seek the shape of the cycle which yields
maximum efficiency $\eta$ under the constraints~\eqref{eq:constraints} on the boundary values of the control
parameters $\lambda$ and $T$~\footnote{
\textcolor{black}{
A similar optimisation problem is often solved in courses on classical thermodynamics to show that maximum efficiency of an equilibrium cycle under the constrains $T(t)\in [T_-, T_+]$ on the bath temperature is the Carnot efficiency. However, in our case, the system can be arbitrarily far form equilibrium.}}. The cycle must run clockwise to secure that $Q_{\mathrm{in}}>Q_{\mathrm{out}}$. Next, we note that maximizing $\eta$ amounts to minimizing the ratio $Q_{\mathrm{out}}/Q_{\mathrm{in}}$. For given boundary values $\sigma_{\pm}$ of $\sigma$, this is obviously achieved by setting $\lambda=\lambda_+$ when $d\sigma>0$ and $\lambda=\lambda_-$ when $d\sigma<0$. In such a case, $Q_{\mathrm{in}} = \lambda_+ \Delta \sigma$, $Q_{\mathrm{out}} = \lambda_- \Delta \sigma$, and the efficiency is given by Eq.~\eqref{eq:opteff}. The increase in the system response $\Delta \sigma = \sigma_+ - \sigma_-$, which can be a complicated functional of the protocol $\{T(t),\lambda(t)\}$, cancelled out. Eq.~\eqref{eq:opteff} is thus valid for arbitrary $\sigma_{\pm}$, and it represents the maximum efficiency of a heat engine based on Hamiltonian~\eqref{eq:general-halm} under the constraints~\eqref{eq:constraints}. The corresponding maximum-efficiency protocol for $\lambda$ forms a rectangle ranging from $\lambda_-$ to $\lambda_+$ in the $\lambda-\sigma$ diagram regardless the cycle duration and dynamical equations of the system. The only constraint on these control parameters is that the cycle runs in the $\lambda-\sigma$ diagram clockwise.} 

\textcolor{black}{
When not driven, a system out of equilibrium relaxes towards the equilibrium state corresponding to the instantaneous values of the fixed control parameters. When the control parameters change periodically, the resulting non-equilibrium state of the system can no longer relax to equilibrium but it lags behind the quasi-static cycle specified by the instantaneous values of the control parameters. In our setting, $\sigma(t)$ lags behind $\sigma^{\rm eq}(t) = \int dx f(x) \exp\{-\lambda(t)f(x)/[k_{\rm B}T(t)]\}/Z(t)$.  
In Sec. B1 of the Appendix, we show that $\sigma^{\rm eq}(t)$ is a monotonously increasing function of $T/\lambda$. Denoting as $t_{+}$ the duration of the $\lambda = \lambda_+$ branch, clockwise cycles with $\Delta \sigma > 0$ are thus obtained for temperature protocols $T(t)$ which obey (i) $\dot{T}(t)\ge 0$ when $\lambda = \lambda_+$, (ii) $\dot{T}(t)\le 0$ when $\lambda = \lambda_-$, and (iii) $T(t_{+}-)/\lambda_+ > T(t_{\rm p}-)/\lambda_-$. The last condition implies that the maximum efficiency~\eqref{eq:opteff} obeys the standard 2nd law inequality $\eta \le 1 - T(t_{\rm p}-)/T(t_{+}-) \le 1 - T_-/T_+$. 
It saturates for the `compression ratio' $\lambda_{-}/\lambda_{+} =
T_{-}/T_{+}$. Even for a  finite cycle time $t_{\rm p}$, output power in this case vanishes. This is because $\sigma^{\rm eq}(t)$ becomes constant, yielding an infinitesimal quasi-static cycle, producing a vanishing output work. In the maximum-efficiency protocol~\eqref{eq:maxEffprot}, we use the specific protocol for $T(t)$ that maximizes the  upper bound on $\eta$. In Sec. B1 of the Appendix, we argue that this temperature protocol also maximizes the output work of the engine regardless $\lambda(t)$ because it yields the largest temperate differences between the bath and the system when they exchange heat.  However, we reiterate that the maximum efficiency~\eqref{eq:opteff} can be achieved for an arbitrary protocol for $T(t)$ that obeys the above conditions (i)-(iii). This freedom in $T(t)$ can be exploited in setups where precise control of the bath (effective) temperature is difficult, such as in active Brownian heat engines~\cite{HolubecActive}.}

\textcolor{black}{The adiabatic branches, where the protocol $\{T(t),\lambda(t)\}$ changes between the boundary values, can be realised using several qualitatively different
approaches~\cite{Holubec2018}. First, one can disconnect the system from the heat bath, which
might be impractical for microscopic engines. Second, one can keep the system in thermal
contact with the bath and vary the control parameters $T$ and $\lambda$ in such
a way that the response $\sigma$ does not change~\cite{Martinez2015}. The
advantage of this approach is that it allows one to circumvent some of the shortcomings of overdamped thermodynamics~\cite{Arold2018}, where the heat fluxes through the momentum degrees of freedom are neglected. Finally, one may realise the adiabatic branches much faster than the relaxation time of the parameter $\sigma$~\cite{blickle2012realization}. In the specific maximum efficiency protocol~\eqref{eq:maxEffprot}, we employ the last possibility, because it minimizes the cycle time $t_{\rm p}$ and thus yields the largest output power $P \equiv W_{\rm out}/t_{\rm p}$. Besides, it allows for a direct comparison with the maximum-efficiency protocols derived for Brownian heat engines under constraints on $\sigma$~\cite{Schmiedl2008}. However, other realisations of the adiabatic branches yield the same maximum efficiency~\eqref{eq:opteff}. Finally, we reiterate that also the choice of the durations $t_{+}$ and $t_{\rm p} - t_{+}$ of the isotherms in the protocol~\eqref{eq:maxEffprot} does not affect the maximum $\eta$.}

\textcolor{black}{
\section{Maximum output power}
If the durations of the isotherms are long enough compared to the relaxation time of the system, i.e., $\Delta \sigma$ is close to its equilibrium value, and the compression ratio $\lambda_-/\lambda_+$ is large, the maximum-efficiency protocol~\eqref{eq:maxEffprot} also yields maximum output work $W_{\rm{out}}$ and power $P$ under the constrained control~\eqref{eq:constraints}. This is our second main result. To prove it, consider the generally unreachable geometric loose upper bound on the output work $\max W_{\rm out} = \Delta \lambda  \max \Delta \sigma^{\rm eq} = (\lambda_+-\lambda_-) [\sigma^{\rm eq}(T_+/\lambda_-)-\sigma^{\rm eq}(T_-/\lambda_+)] $, 
which follows from the broadly valid assumption $\max \Delta \sigma < \max \Delta \sigma^{\rm eq}$ and the insight that $W_{\rm out}$ is given by the area enclosed by the cycle in the $\lambda-\sigma$ diagram. Expanding $\max W_{\rm out}$ in $\Delta \lambda$ yields $\max W_{\rm out} = \Delta \lambda [\sigma^{\rm eq}(T_+/\lambda_+)-\sigma^{\rm eq}(T_-/\lambda_-)] + \mathcal{O}(\Delta \lambda^2)$. 
Up to the leading order in $\Delta \lambda$ and under the condition that the system has relaxed at the ends of the two isotherms to equilibrium, this upper bound is saturated by the protocol~\eqref{eq:maxEffprot}, which completes the proof. We note that: (i) The condition $\Delta \sigma = \Delta \sigma^{\rm eq}$ does not mean that the cycle is slow as the system has to be close to equilibrium at the ends of the two isotherms only and can be arbitrarily far from equilibrium otherwise. (ii) This condition allows one to analytically calculate the whole probability distribution for the output work regardless of additional details of the system dynamics~\cite{Holubec2017, holubec2021fluctuations}. Interestingly, for semi-classical systems, piece-wise constant protocols with two or more branches also maximize output power in the opposite limiting regime, when the cycle time is much shorter than the system relaxation time~\cite{optimalcontrol,erdman2019maximum,cavina2021maximum}.}

\begin{table}[t]
\begin{center}
\begin{tabular}{ c|c|c|c|c } 
&  $\lambda_{\rm pwc}(t)$ & $\lambda_{\mathrm{pwl}}(t)$ & $\lambda_{\mathrm{slow}}(t)$ & $\lambda_{\mathrm{S}}(t)$  \\
 \hline
 \hline
$t<t_+$ & $a$ & $a + bt$ & $\frac{a}{(1+b t)^2}$ & 
$\frac{T_+}{2\sigma_{\rm S}(a,b,t)}-\frac{\sqrt{b}-\sqrt{a}}{\mu t_+\sqrt{\sigma_{\rm S}(a,b,t)}}$ 
\\ 
 \hline
 $t>t_+$ & $c$ & $c + dt$ & $\frac{c}{(1+d t)^2}$ & 
$\frac{T_-}{2\sigma_{\rm S}(a,b,t)}+\frac{\sqrt{b}-\sqrt{a}}{\mu t_-\sqrt{\sigma_{\rm S}(a,b,t)}}$ \\
\end{tabular}
\end{center}
\vspace{-0.3cm}
\caption{\textcolor{black}{Considered classes of protocols with free parameters $\{a, b, c, d\}$ determined by the target of the optimization. The resulting protocols are in general discontinuous at times $t_+$ and $t_{\rm p}$. The protocols $\{\lambda_{\rm pwc}, \lambda_{\rm S}\}$ have two and $\{\lambda_{\rm pwl}, \lambda_{\rm slow}\}$ have four free parameters. The piecewise constant protocol $\lambda_{\rm pwc}(t)$ is a variant of the maximum-efficiency protocol~\eqref{eq:maxEffprot}, where $\lambda(t)$ do not have to reach the boundary values $\lambda_-$ and $\lambda_+$. The piecewise linear protocol $\lambda_{\mathrm{pwl}}(t)$ is perhaps the simplest one that can be implemented in the lab. The protocol $\lambda_{\mathrm{slow}}(t)$ minimizes the irreversible looses during isothermal branches under close-to-equilibrium conditions. Such protocols can be derived for Brownian heat engines with Hamiltonians of
the form $\lambda(t) x^n/n$ (for details, 
see Sec.~D in~the Appendix).
The protocol
$\lambda_{\rm S}(t)$ maximizes both power and efficiency under the constraint that $\sigma(0)=\sigma(t_{\rm p}) \equiv a$ and $\sigma(t_+)\equiv b$~\cite{Schmiedl2008}. The corresponding response $\sigma_{\rm S}(t)$ is given by 
Eq.~(C4) in~the Appendix.
Note that the protocols $\{\lambda_{\rm pwl}, \lambda_{\rm slow}\}$ reduce to $\lambda_{\mathrm{pwc}}(t)$ for $b=d=0$.}}
\label{tab:prot}
\end{table}

\textcolor{black}{Beyond these regimes, $W_{\rm{out}}$ and $P$ strongly depend on all details of the dynamics through $\sigma(t)$ and cycle time $t_{\rm p}$. While $W_{\rm{out}}$ and $P$ are still optimised by the temperature protocol and the choice of fast adiabats in~\eqref{eq:maxEffprot}, optimal protocols for $\lambda(t)$ under the constraints~\eqref{eq:constraints}
are no longer piecewise constant and they have to be identified for each system separately. Similarly as the derivation of maximum-efficiency and maximum-power protocols under constraints on the system state~\cite{Schmiedl2008,Holubec2014,Dechant2017,Zhang2020},
this often involves functional optimisation or extensive numerical work which are both nontrivial tasks.}

\begin{figure*}[t]
\centering
\includegraphics[trim={0cm 2.4cm 4.9cm 0cm},width=0.72\textwidth]{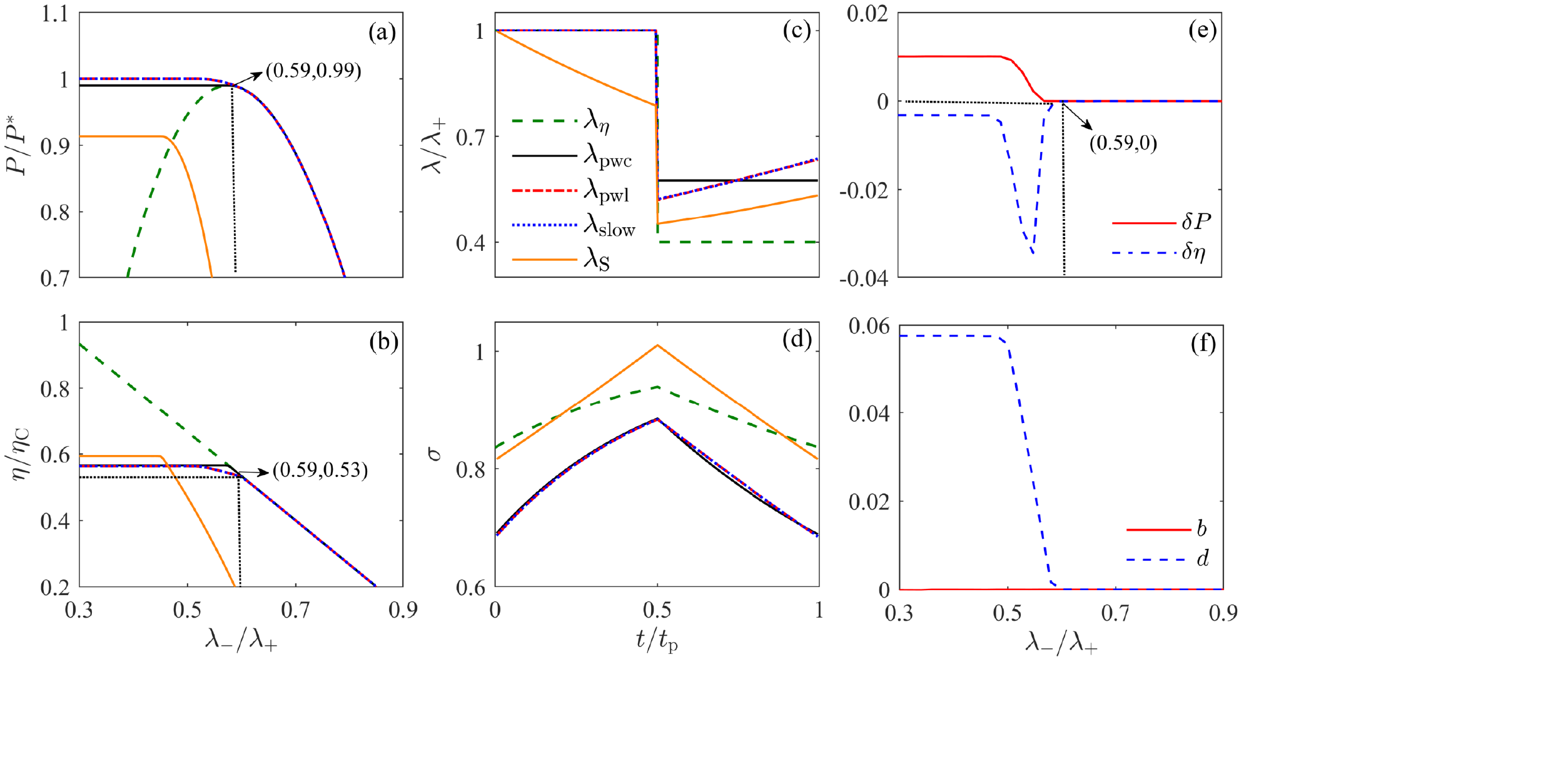}
\caption{\textcolor{black}{Numerical optimization of the output power of the Brownian heat engine illustrates that the maximum-efficiency protocol~\eqref{eq:maxEffprot} also yields maximum power when the compression ratio $\lambda_-/\lambda_+$ is large and the durations $t_+=t_-=1$ of the two isotherms are comparable to the relaxation time $1/(2 \mu \lambda_\pm)$ for $\sigma$.} (a) Powers (in units of the ultimate maximum power $P^*$ for $\lambda_{\rm pwl}$) and (b) the
  corresponding efficiencies obtained using the maximum-efficiency protocol $\lambda_{\eta}$ \eqref{eq:maxEffprot} and the four classes of protocols in Tab.~\ref{tab:prot} optimised for power. For $\lambda_{-}/\lambda_{+} \ge 0.59$ all protocols except for $\lambda_{\rm S}$ coincide. (c) and (d) show the
  protocols and the resulting response for $\lambda_-/\lambda_+=0.4$. \textcolor{black}{(e) the relative differences $\delta X= (X_{\lambda_{\rm pwl}}-X_{\lambda_{\rm pwc}})/X_{\lambda_{\rm pwl}}$ of power ($X=P$) and efficiency ($X=\eta$) for $\lambda_{\rm pwl}$ and $\lambda_{\rm pwc}$. (f) shows the optimal values of paramters $b$ and $d$ for $\lambda_{\rm pwl}$. 
  Parameters used: $k_{\rm B} T_+$=1, $k_{\rm B} T_-$=0.25, $t_+=t_-=1$, $\lambda_+=0.5$, $\mu=1$.}}
\label{fig:compare_uni2}
\end{figure*}

To illustrate the main features of maximum-power protocols for constrained control, we now consider a specific Brownian heat engine based on an overdamped particle diffusing in a controlled harmonic potential. This model describes experimental realizations of microscopic heat engines using optical tweezers~\cite{blickle2012realization, Martinez2016,Martinez2017}. Besides, the corresponding maximum-efficiency and maximum-power protocols under the constrained response is known~\cite{Schmiedl2008}, allowing for a direct comparison with our results. The Hamiltonian~\eqref{eq:general-halm} now reads $H(x,t) = \lambda(t) x^2/2$, with $x$ the position of the
particle. The response of the system $\sigma(t) = \langle x^2/2 \rangle$ is
proportional to the position variance and it obeys the first order differential
equation 
\begin{equation}
  d\sigma(t)/dt = - 2 \mu \lambda(t) \sigma(t) + \mu k_{\rm B} T(t).
  \label{eq:sigmax2}
\end{equation}
It can be solved analytically for an arbitrary mobility $\mu$ of the
particle and any protocol for $\{T(t),\lambda(t)\}$~\cite{holubec2021fluctuations}. In what follows, we measure time,
length and energy in units of $(2\mu\lambda_+)^{-1}$, $\sqrt{k_{\rm B} T_+/(2\lambda_+)}$, and $k_{\rm B}T_+$, respectively.

We ask which protocol for $\lambda$ yields the largest output work and power under the constraints~\eqref{eq:constraints}. Even though the model~\eqref{eq:sigmax2} is exactly solvable, the optimal $\lambda(t)$ has to be found
numerically, e.g., by the method described in Ref.~\cite{Engel2008}. To keep
the optimization transparent, we instead consider a specific set of families of
protocols for the isothermal strokes with free parameters and then numerically
optimize over them. When such classes are chosen suitably, the resulting
suboptimal performance will be close to the global
optimum~\cite{PhysRevE.92.052125,Gingrich2016}. We consider the protocols in Tab.~\ref{tab:prot}. Besides using the protocol for temperature and adiabatic branches form~\eqref{eq:maxEffprot}, we also fix during the optimisation durations of the two isotherms and thus the cycle time.
They can be further optimised once the optimal variation of $\lambda$ is known. 

\textcolor{black}{Main results of the optimisation procedure are summarized in Fig.~\ref{fig:compare_uni2}. (i) With increasing minimum compression ratio $\lambda_-/\lambda_+$ allowed by the constraints~\eqref{eq:constraints}, maximum power for all considered protocols in (a) is first constant and then, at an optimal compression ratio $r^*$, decreases. The decreasing part corresponds to protocols which span between the allowed boundary values, i.e., $\max \lambda(t) = \lambda(0+) = \lambda_+$ and $\min \lambda(t) = \lambda(t_+ +) = \lambda_-$. On the other hand, at the plateau, the boundary values of the protocol are chosen within the bounds~\eqref{eq:constraints} to keep the optimal compression ratio $r^*$. (ii) Values of maximum power obtained for the protocols which have enough free parameters are indistinguishable within our numerical precision.
As the corresponding optimized protocols seem to have minimum possible curvature $\ddot{\lambda}(t)$, we conclude that the maximum-power protocol is piecewise linear.
(iii) Only the protocol $\lambda_{\rm S}$, optimised for constrained response $\sigma$, yields notably smaller power than other protocols. (iv) In agreement with our above discussion, for large enough values
of $\lambda_{-}/\lambda_{+} \ge 0.59$,
the optimized parameters for the protocols $\lambda_{\rm pwl}$, $\lambda_{\rm pwc}$, and $\lambda_{\rm slow}$ are $b=d=0$, $a=\lambda_+$, and $c=\lambda_-$, reducing them to the maximum-efficiency protocol~\eqref{eq:maxEffprot}. (v) The maximum power of the protocols $\lambda_{\eta}$ and $\lambda_{\rm pwl}$ differ just by 1\%. 
In Fig.~D of the Appendix, we further show that for branches durations that further optimize output power, the relative difference is always below 
12\%. This suggests that the maximum-efficiency protocol~\eqref{eq:maxEffprot} can be used to obtain maximum power approximately in situations when $\lambda_{-}$ is a free parameter over which one can further optimise $W_{\rm out}$ and $P$.}

\begin{figure}[t]
\centering
\includegraphics[trim={0cm 0.5cm 0cm 0.23cm},width=0.88\columnwidth]{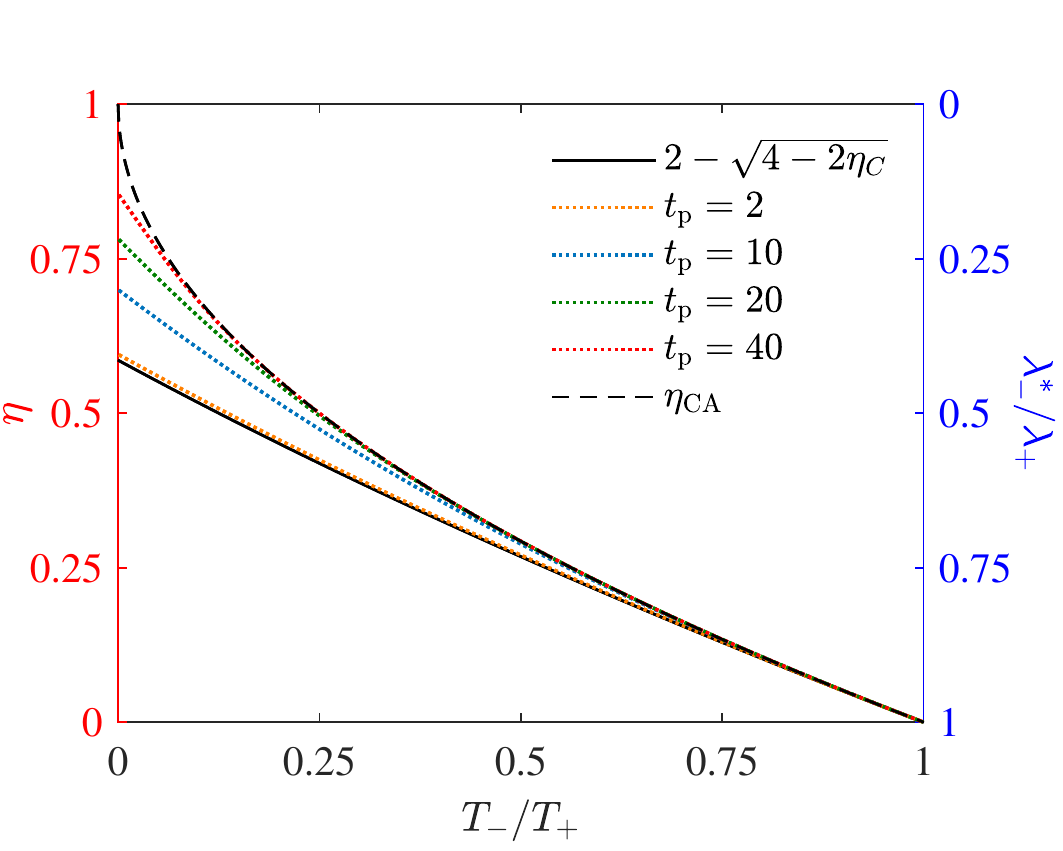}
\caption{Efficiency at maximum output power
($\eta$, red axis) and the corresponding
  optimal compression ratio ($\lambda^*_-/\lambda_+$, blue
  axis) for the maximum-efficiency protocol~\eqref{eq:maxEffprot} as functions
  of the temperature ratio $T_-/T_+$ for six values of cycle
  duration $t_{\rm p}$ (coloured curves) and $t_-= t_+$. The cycle time, $t_{\rm p}$, is measured in units of the system relaxation
  time during the hot isotherm. The corresponding ratio during the cold
  isotherm reads $t_{\rm p}\, \lambda^*_-/\lambda_+$. We used the same parameters as in Fig. \ref{fig:compare_uni2}.
  }
  \label{fig:piece}
\end{figure}

Such optimisation seems a natural task for various
experimental platforms where the main limitation is on the maximum strength of the potential. The maximum power regime of the maximum-efficiency protocol can be, to a large extend, investigated analytically.
First,  assuming again that durations of the isotherms
are long enough that the system is close to equilibrium at times $t_+$ and $t_{\rm p}$,
we have
$W_{\rm{out}}= (\lambda_{+} -
\lambda_{-})[\sigma^{\rm eq}(T_+/\lambda_+)-\sigma^{\rm eq}(T_-/\lambda_-)]$. For $f(x) = \lvert{x}\rvert^n$ in Eq.~\eqref{eq:general-halm}, we then find that the optimal compression ratio is
$\lambda_{-}/\lambda_{+} =
\sqrt{T_{-}/T_{+}}$, which leads to the output work
$W_{\mathrm{out}} = k_{\mathrm{B}} T_{+} (2\eta_{\mathrm{CA}} -
\eta_{\mathrm{C}})/n$ and Curzon-Ahlborn efficiency $\eta = \eta_{\mathrm{CA}}
= 1 - \sqrt{T_{-}/T_{+}}$ (see 
Sec. B1 of~the Appendix for details). For other than power-law Hamiltonians, the
efficiency at maximum power can differ from $\eta_{\mathrm{CA}}$ but it can still be determined numerically regardless details of dynamical equations for the system (for details, see 
Sec. B1 in~the Appendix). Relaxing the assumption of slow (but not quasi-static) isotherms, the optimization of
$W_{\mathrm{out}}$ with respect to $\lambda_{-}$ requires
specification of the dynamics.  In Fig.~\ref{fig:piece}, we show that the efficiency at maximum power of the Brownian heat engine described by Eq.~\eqref{eq:sigmax2} and driven by the maximum-efficiency protocol~\eqref{eq:maxEffprot} is
bounded between the Curzon-Ahlborn efficiency, achieved for slow isotherms, and
the efficiency $2 - \sqrt{4 - 2\eta_{\mathrm{C}}} < \eta_{\mathrm{CA}}$,
reached in the limit of infinitely fast cycles (see 
Sec. B in~the Appendix for details).



\section{Conclusion}
The main strength of the presented derivations of the maximum-efficiency and maximum-power protocols under constrained control is their simplicity and unprecedented generality. A possible extension of these results to more complicated Hamiltonians is sketched in
Sec.~A of
the Appendix. However, more general extensions remain to be explored in
future work. \textcolor{black}{Nevertheless, the validity of our results for Brownian heat engines is already of experimental relevance as they are often realized using optical tweezers with strict bounds on the trap stiffness $\lambda$ -- too small $\lambda$ leads to losing the Brownian particle while too large $\lambda$ can induce its overheating and even evaporation. At the same time, the achievable trap stiffnesses are well above $10^{-6}$ N/m~\cite{Holubec2018}. For a spherical Brownian particle with the radius of $10^{-6}$ m in water, the Stokes law predicts the mobility of $\mu\approx 0.5\times 10^8$ m/Ns, leading to the relaxation time $1/(2\mu \lambda)$ of the response $\sigma$ on the order of $10^{-2}$ s. The assumption that the durations of the isotherms are comparable to the response relaxation time, used in our derivation of the maximum-power protocol, is thus, in this setup, natural.} Besides, we believe that extensions of our results can find applications in more involved optimization tasks, e.g., performed using machine
learning algorithms~\cite{erdman2021identifying, khait2021optimal} or geometric
methods \cite{dechant2019thermodynamic,Vu2021}, as well as in quantum setups \cite{abiuso2020geometric,PRXQuantum.3.010323,Pancotti2020}.


\textit{Acknowledgements.}
ZY is grateful for the sponsorship of China Scholarship Council (CSC) under
Grant No. 201906310136.   
FC gratefully acknowledges funding from the Fundational Questions Institute Fund (FQXi-IAF19-01).
VH gratefully acknowledges support by the Humboldt
foundation and by the Czech Science Foundation (project No. 20-02955J). 
PA is supported by “la Caixa” Foundation (ID 100010434, Grant No. LCF/BQ/DI19/11730023), and by the Government of Spain (FIS2020-TRANQI and Severo Ochoa CEX2019-000910-S), Fundacio Cellex, Fundacio Mir-Puig, Generalitat de Catalunya (CERCA, AGAUR SGR 1381.
MPL acknowledges financial support from the Swiss National Science Foundations (Ambizione grant PZ00P2-186067).
JA acknowledges funding from the Engineering and Physical Sciences Research Council (EPSRC) (EP/R045577/1) and thanks the Royal Society for support.


\bibliography{Refs_main}

\appendix


\section{Maximum-efficiency protocol for multi-term Hamiltonians}
\label{supp-sm-general}

Consider a heat engine with a working fluid described by the Hamiltonian 
\begin{equation}
H(x,t) = \sum_i \lambda_i(t) f_i(x)
\label{eq:Hfuj}
\end{equation}
with control parameters $\lambda_i(t)$, $i=1,\dots,N$. As in the main text, we now aim to derive the finite-time protocol for the constrained control parameters, $\lambda_i(t) \in (\lambda_i^{-}, \lambda_i^{+})$, which would yield maximum efficiency of the engine. It will turn out that if the compression ratios $\lambda_i^{-}/ \lambda_i^{+}$ for all the control parameters equal, the geometric argument from the main text still applies.

The heat increment is for the Hamiltonian~\eqref{eq:Hfuj} given by \dj $Q = \sum_i \lambda_i(t) d\sigma_i(t)$ with the response functions $\sigma_i(t) = \langle f_i(x) \rangle$. For arbitrary fixed maximum changes $\Delta \sigma_i$ in the response functions during the cycle, geometric upper and lower bounds on $Q_{\rm in}$ and $Q_{\rm out}$ and thus on efficiency are achieved by clockwise rectangular cycles in the individual $\lambda_i-\sigma_i$ diagrams. These hypothetical cycles yield the following geometric upper bound on efficiency:
\begin{equation}
  \eta = 1 - \frac{Q_{\rm out}}{Q_{\rm in}} \le 1 - \frac{\sum_i \Delta \sigma_i \lambda_i^{-}}{\sum_i \Delta \sigma_i \lambda_i^{+}}.
  \label{eq:effUB}
\end{equation}
We use the term `geometric' to stress that this bound follows from the analysis of the cycle in the $\lambda-\sigma$ diagram, without considering the relation between the protocol $[\lambda_1(t),\dots,\lambda_N(t)]$ and the response $[\sigma_1(t),\dots,\sigma_N(t)]$ imposed by dynamical equations of the working fluid.
This means that the given set of $\Delta \sigma_i$ might not be achievable by the piecewise constant protocol and thus the bound in~\eqref{eq:effUB} is loose. Furthermore, we seek an optimal protocol constrained just by the conditions on $\lambda_i$ and the upper bound in~\eqref{eq:effUB} in general strongly depends on the fixed values of $\Delta \sigma_i$. For the single-term Hamiltonian $H(x,t) = \lambda(t) f(x)$ used in the main text, this has not been an issue because then $\Delta \sigma$ in the nominator and denominator in \eqref{eq:effUB} cancel out and the upper bound becomes independent of the details of the dynamics. The optimal protocol for efficiency is then the piecewise constant protocol for $\lambda(t)$ because it saturates the geometric upper bound. To sum up, the bound in~\eqref{eq:effUB} allows to derive the maximum-efficiency protocol only if it happens to be independent of $\Delta \sigma_i$. In the opposite case, the optimal protocol cannot be determined without considering the dynamical equations and performing the corresponding functional optimisation. 

Let us now investigate  when the upper bound in~\eqref{eq:effUB} becomes independent of the system response, $\Delta \sigma_i$. Defining the set of `probabilities' $p_i =\Delta\sigma_i \lambda_i^{+}/\sum_i \Delta \sigma_i \lambda_i^{+}$, the ratio in the upper bound in~\eqref{eq:effUB} can be rewritten as the average 
\begin{equation}
 \frac{\sum_i \Delta \sigma_i \lambda_i^{-}}{\sum_i \Delta \sigma_i \lambda_i^{+}} = \sum_i p_i \frac{\lambda_i^{-}}{\lambda_i^{+}}.
 \label{eq:prob}
\end{equation}
This expression becomes independent of $\sigma_i$ only if all the compression ratios $\lambda_i^{-}/ \lambda_i^{+}$ are equal. In such a case, the maximum-efficiency protocol is thus a piecewise constant protocol for each of $\lambda_i$ and yields the efficiency
\begin{equation}
    \eta = 1 - \lambda_i^{-}/ \lambda_i^{+}.
    \label{eq:bound_loose}
\end{equation}
Besides this result, the probabilistic interpretation~\eqref{eq:prob} of the upper bound in~\eqref{eq:effUB} also yields the dynamics independent (but in general loose) \emph{upper bound} on efficiency,
\begin{equation}
    \eta \le 1 - \min_i \frac{\lambda_i^{-}}{\lambda_i^{+}}.
    \label{eq:upp_bou}
\end{equation}
To close this section, we note that a piecewise constant protocol for $\lambda_i$ will always yield the efficiency
$
     1 - (\sum_i \Delta \sigma_i \lambda_i^{-})/(\sum_i \Delta \sigma_i \lambda_i^{+}),
$
with values of $\Delta \sigma_i$ induced by the dynamical equations of the system. Within the class of piecewise constant protocols, the upper bound~\eqref{eq:upp_bou} is then tight if the constraints on all the control parameters $\lambda_i$ allow to achieve the minimum compression ratio
$\min_i \frac{\lambda_i^{-}}{\lambda_i^{+}}$. Furthermore, for such protocols, Eq.~\eqref{eq:prob} also implies the \emph{lower bound} on the efficiency
\begin{equation}
    \eta \ge 1 - \max_i \frac{\lambda_i^{-}}{\lambda_i^{+}},
    \label{eq:low_bou}
\end{equation}
which is always tight.

\section{Properties of maximum-efficiency protocol}
\label{sec:maxeff}

In this section, we provide further details concerning the maximum-efficiency protocol for the Hamiltonian, $H(x,t)=\lambda(t)f(x)$, discussed in the main text. \textcolor{black}{First, we argue that the maximum-efficiency protocol that yields maximum output work for the given piecewise constant $\lambda(t)$
requires piecewise constant variation of temperature.} Then, we investigate output power of the maximum-efficiency protocol as a function of the lower bound on the control parameter $\lambda(t)$. 

\subsection{Temperature protocol}
\label{suppl-mat-setting-opt}

In the main text, we have shown that the maximum-efficiency protocol for the control parameter $\lambda(t)$ is piecewise constant and the corresponding efficiency $\eta = 1 - \lambda_{-}/\lambda_{+}$. The only condition on the temperature protocol was that the cycle is performed clockwise in the $\lambda-\sigma$ diagram. Nevertheless, in order to allow the engine to operate at Carnot efficiency and to maximize its output work, we have chosen the protocol 
\begin{equation}
\{T(t), \lambda(t)\}_{\eta}=
\begin{cases}
\{T_+, \lambda_+\}, &0<t<t_+\\
\{T_-,\lambda_-\}, &t_+<t<t_{\rm p}
\end{cases}.
\label{expression-sm-piece-protocol}
\end{equation}

\textcolor{black}{For this choice of $T(t)$,
the working medium of the engine operates with the largest possible temperature gradient during the whole cycle. This maximizes the heat flux through the engine, which can be utilised to yield the maximum amount of work $W_{\rm out} = \eta Q_{\rm in}$. Besides, the engine efficiency $\eta$ is also known to increase with the bath temperature difference [see also Fig. \ref{fig:diff} (c)-(f)].}

\textcolor{black}{Let us now provide an alternative and more technical argument that the choice of $T(t)$ in Eq.~\eqref{expression-sm-piece-protocol} maximizes the output work. We restrict this argument to the maximum-efficiency protocol for $\lambda$ in Eq.~\eqref{expression-sm-piece-protocol}. However, generalizations to other protocols are straightforward. The main idea is that connecting the system to the hottest possible bath when $\dot{\sigma}>0$ and to the coldest possible bath when $\dot{\sigma}<0$ maximizes the extent of the cycle in the $\sigma$ direction in the $\sigma-\lambda$ diagram and thus also $W_{\rm out}$.}

\textcolor{black}{
For the protocol~\eqref{expression-sm-piece-protocol}, the output work is given by
\begin{equation}
    W_{\rm out} = \Delta \lambda \Delta \sigma,
    \label{uni2-diss-sm-output-work}
\end{equation}
with $\Delta \lambda = \lambda_+ - \lambda_-$ and the maximum change in the response parameter during the cycle $\Delta \sigma = \sigma_+ - \sigma_-$. To maximize $W_{\rm out}$, we thus need to maximize $\Delta \sigma$. To this end, it is reasonable to assume that 
\begin{equation}
\Delta \sigma \le \Delta \sigma^{\rm eq}, 
\label{suppl-mat-bound-relation}
\end{equation}
where $\Delta \sigma^{\rm eq}=\max\sigma^{\rm eq}-\min\sigma^{\rm eq}$ is the maximum change in the response parameter $\sigma$ during the cycle with isochoric branches (constant $\lambda$) longer than the system relaxation time. This assumption is in particular valid for arbitrary overdamped dynamics, where $\sigma$ always converges to its equilibrium value ($k_{\rm B}$ denotes the Boltzmann constant)
\begin{equation}
\sigma^{\rm eq}(t) = \sum_x f(x) \frac{\exp\left\{- \lambda(t)f(x)/[k_{\rm B} T(t)]\right\}}{\sum_x \exp\left\{-\lambda(t)f(x)/[k_{\rm B} T(t)]\right\}},
\label{eq:EQstate}
\end{equation}
corresponding to the instantaneous values of the control parameters $\{T(t),\lambda(t)\}$. Noticing that $\sigma^{\rm eq}(t) = U(t)/\lambda(t)$, where $U(t) = \left< H(x,t) \right>$ is the thermodynamic internal energy of the system, the positivity of heat capacity
\begin{equation}
C_{\rm v} = \frac{\partial{U}}{\partial{T}} =  \frac{\partial{ \sigma^{\rm eq}}}{\partial (T/\lambda)}  > 0
\label{eq:Cv}
\end{equation}
implies that $\sigma^{\rm eq}$ is a monotonously increasing function of the ratio $T/\lambda$.}

\textcolor{black}{From Fig.~1 in the main text, it follows that $\max\sigma^{\rm eq}$ and $\min\sigma^{\rm eq}$ are the values of $\sigma^{\rm eq}$ at the ends of the isochores with $\lambda = \lambda_{+}$ and $\lambda = \lambda_{-}$, respectively. The upper bound on $\Delta \sigma$ is thus given by
\begin{equation}
\max\sigma^{\rm eq}-\min\sigma^{\rm eq} = \sigma^{\rm eq}(T_{+}/\lambda_{+}) - \sigma^{\rm eq}(T_{-}/\lambda_{-}). 
\label{sm-suppl-material-max-fir}
\end{equation}
It is attained for slow isochores when $T=T_{+}$ for $\lambda = \lambda_{+}$ and $T=T_{-}$ for $\lambda = \lambda_{-}$. As long as $\dot{\sigma}^{\rm eq} > 0$ for $\lambda = \lambda_{+}$ and $\dot{\sigma}^{\rm eq} < 0$ for $\lambda = \lambda_{-}$ (so that the used definitions of input and output heat hold), details of the temperature protocol during the isochores in this limit do not alter the value of $\Delta \sigma^{\rm eq}$ and thus $W_{\rm out} = \Delta \lambda \Delta \sigma^{\rm eq}$. However, these details become important for finite-time cycles.}

\textcolor{black}{
A typical dynamical equation for an overdamped degree of freedom has the form
\begin{equation}
\dot{\sigma}(t) = t_{\rm R}^{-1} [\sigma^{\rm eq}-\sigma(t)].
\label{eq:XXX}
\end{equation} 
For constant values of control parameters $T(t)$ and $\lambda(t)$, which enter the relaxation time $t_{\rm R}$ and the equilibrium state $\sigma^{\rm eq}(t)$ defined in Eq.~\eqref{eq:EQstate}, this equation describes an exponential relaxation of $\sigma$ to $\sigma^{\rm eq}$ (for a specific example, see Sec.~\ref{SI-section-short-isotherm}). For a cyclic variation of the control parameters, $\sigma$ follows $\sigma^{\rm eq}$ like a donkey chasing a carrot \cite{holubec2014non}. As a result,
$\sigma \le \sigma^{\rm eq}$ and $\dot \sigma \ge 0$ for $\lambda = \lambda_{+}$, when $\sigma^{\rm eq}$ increases to $\max\sigma^{\rm eq}$, and $\sigma \ge \sigma^{\rm eq}$ and $\dot \sigma \le 0$ for $\lambda = \lambda_{-}$, when $\sigma^{\rm eq}$ decreases to $\min\sigma^{\rm eq}$. The change in the response $\Delta \sigma = \int_0^{t_{+}} \dot{\sigma} dt = - \int_{t_{+}}^{t_{\rm p}} \dot{\sigma} dt$ and thus it can be maximized by maximizing (minimizing) the instantaneous rate of change of the response, $\dot{\sigma}$, during the first (second) isochore.  From Eq.~\eqref{eq:XXX}, it follows that this is achieved by setting $\sigma^{\rm eq} = \max\sigma^{\rm eq}$ during the fist isochore and $\sigma^{\rm eq} = \min\sigma^{\rm eq}$ during the second one.
Altogether, this suggests that the piecewise constant temperature protocol in Eq.~\eqref{expression-sm-piece-protocol} yields maximum $\Delta \sigma$ and thus
output work $W_{\rm out}$ \eqref{uni2-diss-sm-output-work} for arbitrary cycle duration.}

\begin{figure}[t]
\centering
\includegraphics[trim={0cm 0.7cm 0.5cm 0cm},width=0.8\columnwidth]{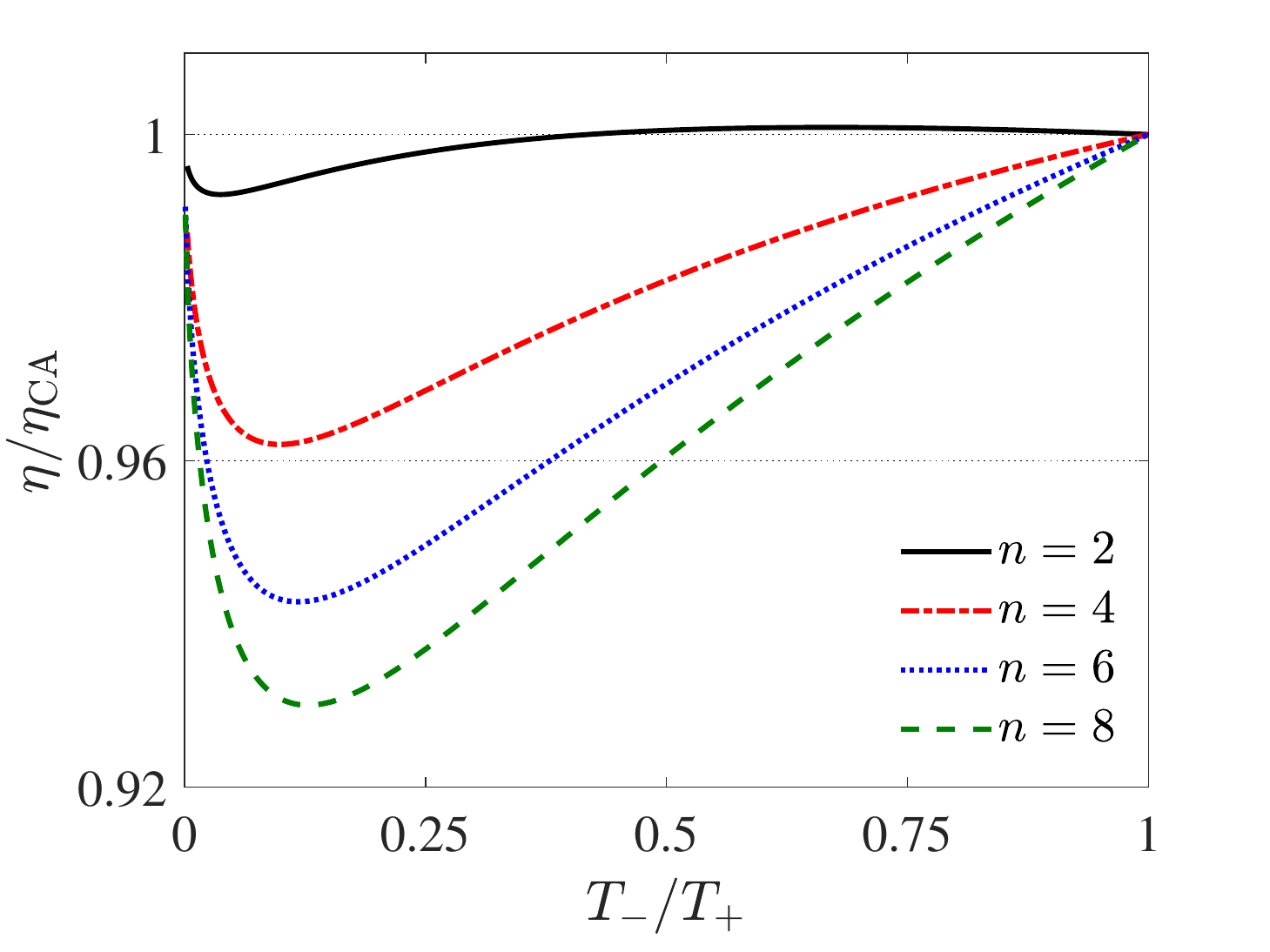}
\caption{Efficiency at maximum output work obtained using the Hamiltonian $H = \lambda(t) (|x|^n/n - \ln|x|)$ as a function of $T_-/T_+$.  Parameters used: $k_{\rm B} T_+=1$ and $\lambda_+=0.5$.}
\label{fig:asymmetri}
\end{figure}

\subsection{Maximum output work of maximum-efficiency protocol}
\label{sm-sec-piecewise}

Let us now turn to the task of maximizing the output work $W_{\rm out} = (\lambda_{+} - \lambda_{-}) \Delta \sigma$ with respect to $\lambda_{-}$. 
Analytical results can be obtained in the limits of slow and fast isotherms. 

\subsubsection{Slow isotherms}
\label{suppl-mat-slow-sot}

When the duration of the isotherms is
longer than the relaxation time of the response $\sigma$, one can approximate $\sigma_+$ and $\sigma_-$ in $\Delta \sigma$ by their equilibrium values. Using Eq.~\eqref{uni2-diss-sm-output-work}, the output work then reads 
\begin{equation}
W_{\rm out}= \Delta\lambda\Delta \sigma^{\rm eq}.
\label{uni2-diss-sm-output-work-eq-d}
\end{equation}
Equation~\eqref{eq:Cv} implies that the partial derivative of $\sigma^{\rm eq}$ with respect to the control parameter $\lambda$ ($T$ is constant) is given by
\begin{equation}
\frac{\partial}{\partial \lambda}\sigma^{\rm eq}(T/\lambda) =  - \frac{T}{\lambda^2}C_{\rm v}.
\label{SI-first-sigma-eq-lam}
\end{equation}
The condition on the extreme of $W_{\rm out}$ \eqref{uni2-diss-sm-output-work-eq-d} with respect to $\lambda_-$ thus reads:
\begin{equation}
\begin{split}
\frac{\partial{W_{\rm out}}}{\partial \lambda_-}
&= (\lambda_+-\lambda_-) \frac{T_-}{\lambda_-^2}C_{\rm v}(T_-/\lambda_-)\\
&- \frac{U(T_+/\lambda_+)}{\lambda_+} + \frac{U(T_-/\lambda_-)}{\lambda_-}= 0,    
\end{split} 
\label{eq:kmin}
\end{equation}
where we additionally used the relation $\sigma^{\rm eq} = U/\lambda$ between $\sigma^{\rm eq}$ and the internal energy $U$.

For power law Hamiltonians of the form $H = \lambda |x|^n/n$ where $C_{\rm v} = k_{\rm B}/n$ and $U=k_{\rm B} T/n$, this equation can be solved explicitly. The resulting optimal compression ratio is given by $\lambda_-/\lambda_+ = \sqrt{T_-/T_+}$. The corresponding efficiency at the maximum output work is given by the Curzohn-Ahlborn efficiency,
\begin{equation}
    \eta = 1 - \frac{\lambda_-}{\lambda_+} = 1-\sqrt{\frac{T_-}{T_+}} \equiv \eta_{\rm CA},
    \label{uni2-diss-eff}
\end{equation}
and the maximum output work is (Carnot efficiency $\eta_{\rm C} = 1 - T_-/{T_+}$)
\begin{equation}
    W_{\rm out}=\frac{k_{\rm B} T_+}{n}(2\eta_{\rm CA}-\eta_{\rm C}).
\end{equation}

Let us now consider the asymmetric Hamiltonian $H = \lambda(t) (|x|^n/n - \ln|x|)$. In this case, the internal energy and heat capacity are given by
\begin{eqnarray}
U&=&\frac{k_{\rm B}T+\lambda\left[1+\ln{\frac{\lambda}{nk_{\rm B}T}}-\psi^{(0)}\left(\frac{\lambda+k_{\rm B}T}{nk_{\rm B}T}\right)\right]}{n},\\
C_{\rm v}&=&\frac{nk_{\rm B}T(k_{\rm B}T-\lambda)+\lambda^2\psi^{(1)}\left(\frac{\lambda+k_{\rm B}T}{nk_{\rm B}T}\right)}{n^2k^2_{\rm B}T^2}, 
\end{eqnarray}
where $\psi^{(m)}(z)$ denotes the polygamma function of order $m$. In this case, Eq.~\eqref{eq:kmin} is transcendental and we solved it numerically. In Fig. \ref{fig:asymmetri}, we show the resulting efficiency at the maximum output work as a function of $T_-/T_+$. Even though the resulting efficiency is still close to $\eta_{\rm CA}$, it can be both slightly larger and smaller than that.

\subsubsection{Fast isotherms}
\label{SI-section-short-isotherm}

Let us now assume that the duration of the isothermal branches are much shorter than the system relaxation time. In such a case, the work optimisation cannot be done without specifying the dynamical equation for the response $\sigma$. To this end, we assume that it obeys the overdamped equation 
\begin{equation}
\dot \sigma(t) = t_{\rm R}^{-1} [\sigma^{\rm eq} - \sigma(t)]    
\label{eq:gen_over}
\end{equation}
with the equilibrium value $\sigma^{\rm eq}$ and relaxation time $t_R$ determined by the values of the control parameters $\{T(t),\lambda(t)\}$ at time $t$. The most prominent examples of systems described by this formula is a two-level system~\cite{erdman2019maximum} and an overdamped particle trapped in a harmonic potential~\cite{Schmiedl2008}.

Solving Eq.~\eqref{eq:gen_over} for the maximum-efficiency protocol~\eqref{expression-sm-piece-protocol}, we find that
\begin{equation}
\sigma(t) =
\begin{cases}
\sigma_0 {\rm e}^{-\frac{t}{t_{R}^{+}}} + \sigma_{\rm eq}^{+} (1- {\rm e}^{-\frac{t}{t_{R}^{+}}}), &0<t<t_+\\
\sigma_1 {\rm e}^{-\frac{t-t_{+}}{t_{R}^{-}}} + \sigma_{\rm eq}^{-} (1- {\rm e}^{-\frac{t-t_{+}}{t_{R}^{-}}}), &t_+<t<t_{\rm p}
\end{cases},
\label{expression-sm-piece-protocol-1}
\end{equation}
where $\sigma_0 \equiv \sigma(0)$ and $\sigma_1 \equiv \sigma(t_{+})$ are determined by the condition that $\sigma(t)$ must be a continuous function of time. The variables corresponding to the first (second) isotherm are denoted by max (min). It turns out that 
\begin{multline}
\Delta \sigma=\sigma_{+} - \sigma_{-} = \sigma_1 - \sigma_0 =\\
\Delta\sigma^{\rm eq} 
\frac{\sinh\left(t_{+}/t_{\rm R}^{+}\right) \sinh\left(t_{-}/t_{\rm R}^{-}\right)}{\sinh\left(t_{+}/t_{\rm R}^{+}+t_{-}/t_{\rm R}^{-}\right)} \le \Delta\sigma^{\rm eq} .
\end{multline}
Substituting this result into the expression for the output work \eqref{uni2-diss-sm-output-work} and expanding the result up to the leading order in the ratios of duration of the individual isotherms to the corresponding relaxation times, $t_{+}/t_{\rm R}^{+}$ and $t_{-}/t_{\rm R}^{-}$, we find that
\begin{equation}
W_{\rm out}=\Delta\lambda\Delta\sigma^{\rm eq}  \frac{\frac{t_{+} t_{-}}{t_{\rm R}^{+}t_{\rm R}^{-}}}{\frac{t_{+}}{t_{\rm R}^{+}}+\frac{t_{-}}{t_{\rm R}^{-}}}.
\label{SI-general-work-out-tr}
\end{equation}
To maximize the output work, we need to choose a specific model to determine the dependence of the equilibrium values of response and relaxation times on the control parameters. To this end, we consider the paradigmatic model of stochastic thermodynamics, $\dot{\sigma}(t) = -2 \, \mu \, \lambda(t) \, \sigma(t) + \, \mu \, k_{\rm B} T$, describing an overdamped Brownian particle with mobility $\mu$ in a harmonic trap~\cite{Schmiedl2008, PhysRevLett.98.108301}. In this case, $\sigma^{\rm eq} = T/(2\mu \lambda)$ and $t_R = 1/(2\mu \lambda)$, and the maximum output work \eqref{SI-general-work-out-tr} is produced for 
\textcolor{black}{
\begin{equation}
\frac{\lambda_{-}}{\lambda_{+}}
=\sqrt{(\alpha+1)\left(\alpha+1-\eta_{\rm C}\right)}-\alpha,
\end{equation}
where $\alpha \equiv t_{+}/t_{-}$. The corresponding efficiency reads
\begin{equation}
\eta=1-\frac{\lambda_-}{\lambda_+}=\alpha+1-\sqrt{(\alpha+1)(\alpha+1-\eta_{\rm C})}, 
\label{SI-S26-EMP}
\end{equation}
which reduces to $\eta_{\rm CA}$ for $\alpha\to 0$ and $\eta_{\rm C}/2$  for $\alpha\to\infty$.
Assuming that $\alpha = 1$ ($t_{+}=t_{-}$), Eq. \eqref{SI-S26-EMP} is given by the formula
\begin{equation}
\eta=2-\sqrt{4-2\eta_{\rm C}}
=\frac{\eta_{\rm C}}{2}+\frac{\eta^2_{\rm C}}{16}+\mathcal{O}(\eta^3_{\rm C})
\label{SI-fast-alpha-eff}
\end{equation}
used in the main text.} 
The corresponding expansion for the Curzohn-Ahlborn efficiency, $\eta_{\rm CA}\approx\frac{\eta_{\rm C}}{2}+\frac{2\eta_{\rm C}^2}{16}$, has identical linear and twice larger quadratic term.
\textcolor{black}{Last but not least, with respect to $\alpha$, the output power $W_{\rm out}/t_{\rm p}$ using Eq. \eqref{SI-general-work-out-tr}  develops a peak at $\alpha=\alpha^*=\sqrt{\frac{\lambda_-}{\lambda_+}}<1$. 
This also contradicts the situation with constrained $\sigma$, where maximum power is attained when the durations of the isotherms are equal ($\alpha=\alpha^*=1$) \cite{Schmiedl2008}.
}

\section{Numerical results}
\label{SI-comarison-control-response}

In this section, we give further details concerning the numerical optimisation of output work (and power) described in the main text. All the numerical results were obtained using the model 
\cite{Schmiedl2008, PhysRevLett.98.108301}
\be \label{eq:overdamped-har-sm}
  \dot{\sigma}(t) = -2 \, \mu \, \lambda(t) \, \sigma(t) + \, \mu \, k_{\rm B} T
\ee
for an overdamped Brownian particle with mobility $\mu$ in a harmonic trap with stiffness $\lambda(t)$, which well describes experiments with optical tweezers. This equation is exactly solvable for arbitrary protocol $\{T(t),\lambda(t)\}$. However, the solutions can involve exponentials of very large or small numbers, which can lead to numerical instabilities inducing large looses of precision, and thus they have to be treated with care. To secure that our solutions to Eq.~\eqref{eq:overdamped-har-sm} are always enough precise, we have solved it in our analysis also numerically.

\textcolor{black}{
For $\lambda(t)$, we considered the maximum-efficiency protocol \eqref{expression-sm-piece-protocol} 
and the four classes of protocols in Tab. I in the main text. For $T(t)$, we always used the protocol in Eq. \eqref{expression-sm-piece-protocol}.} For these protocols, we optimised the output power and efficiency 
\begin{equation}
P=\frac{Q_{\rm in}-Q_{\rm out}}{t_{\rm p}},\quad \eta=1-\frac{Q_{\rm out}}{Q_{\rm in}},    
\end{equation}
with the input heat $Q_{\rm in}=\int_{0}^{t_+}dt\dot{\sigma}(t)\lambda(t)$ and the output heat $Q_{\rm out}=-\int_{t_+}^{t_{\rm p}}dt\dot{\sigma}(t)\lambda(t)$, \
\textcolor{black}{
as functions of the parameters $\{a, b, c, d\}$ (see Tab. I in the main text) under different constraints.}

\textcolor{black}{In Sec. \ref{SI-constrained-response-I}, we verify that the protocol $\lambda_{\rm S}$ from Ref.~\cite{Schmiedl2008} yields both maximum power and efficiency under constrains on the response $\sigma$. In Sec. \ref{SI-constrained-control-I}, we provide numerical results illustrating that the protocol \eqref{expression-sm-piece-protocol} yields maximum efficiency under constrains on the response $\lambda$. Furthermore, we
present additional evidence that the maximum-efficiency protocol \eqref{expression-sm-piece-protocol} optimized with respect to $\lambda_-$ can be used to approximately obtain also the maximum power.
}

\subsection{Constrained response}
\label{SI-constrained-response-I}

\begin{figure}[t]
\centering
\includegraphics[trim={0cm 6.6cm 2.1cm 0cm},width=0.99\columnwidth]{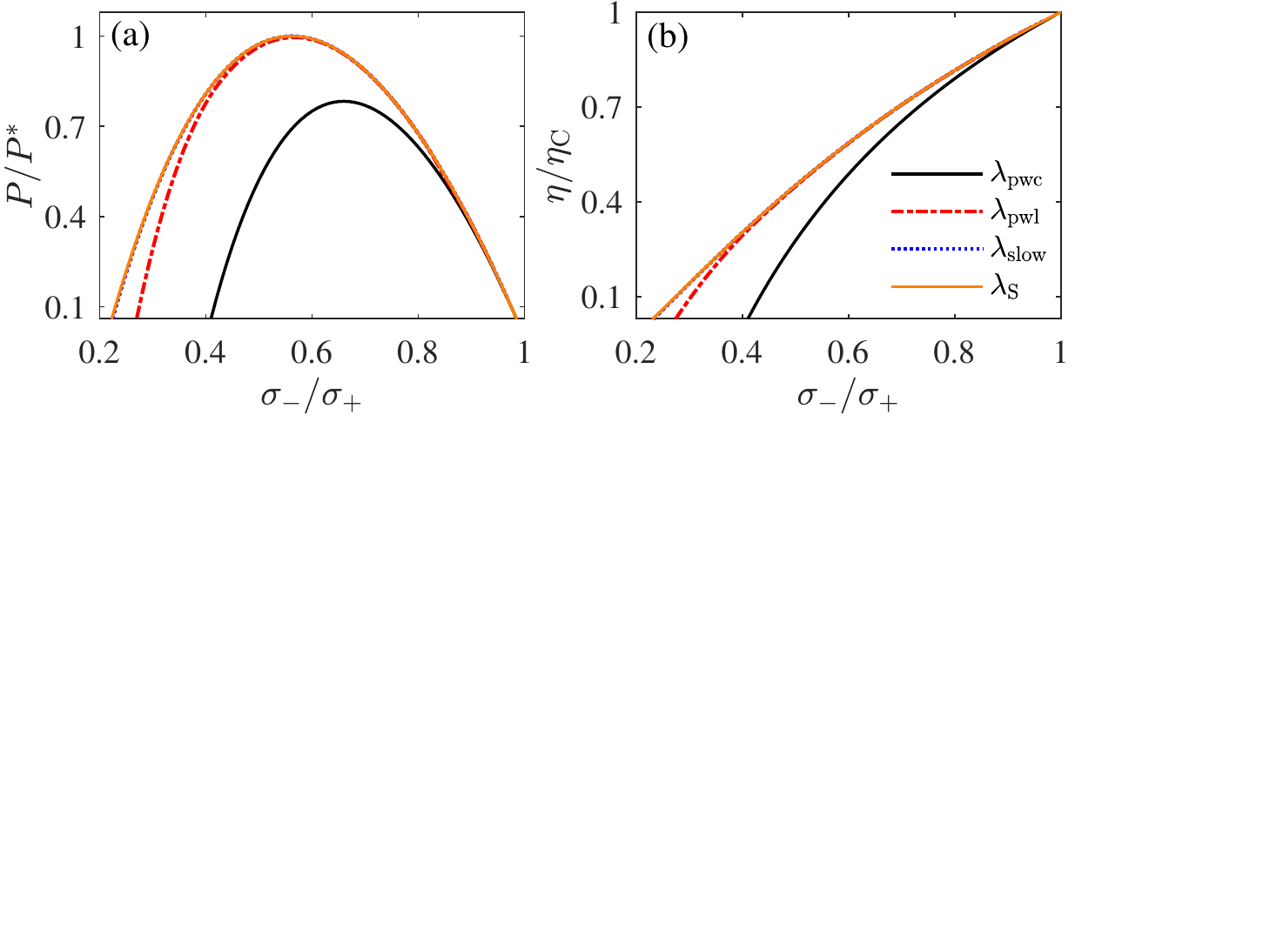}
\caption{Optimal performance for fixed boundary values of the response\textcolor{black}{: $\sigma_-\equiv\sigma(0)=\sigma(t_{\rm p})$ and $\sigma_+\equiv\sigma(t_+)=0.5$}. (a) Maximum power (in units of the ultimate maximum power $P^*$ \textcolor{black}{for $\lambda_{\rm pwl}$}) and (b) maximum efficiency as functions of $\sigma_-/\sigma_+$. 
Lines corresponding to $\lambda_{\rm S}$ (cyan dashed) and $\lambda_{\rm slow}$ (blue dotted) perfectly overlap.
\textcolor{black}{The maximum-efficiency protocol \eqref{expression-sm-piece-protocol} and the piecewise constant protocol $\lambda_{\rm pwc}$ are in this case equal. We used the same parameters as in Fig. 2 in the main text.}
}
\label{SI-uni1}
\end{figure}

\textcolor{black}{
When the values of the response (position variance) $\sigma$ at the ends of the two isotherms are fixed, i.e., $\sigma_-\equiv\sigma(0)=\sigma(t_{\rm p})$ and $\sigma_+\equiv\sigma(t_+)$, the protocol which yields both maximum efficiency and power reads~\cite{Schmiedl2008}}
\begin{equation}
\lambda_{\rm S}=
\begin{cases}
\frac{T_+}{2\sigma_{\rm S}}-\frac{\sqrt{\sigma_{+}}-\sqrt{\sigma_{-}}}{\mu t_+\sqrt{\sigma_{\rm S}}},~~ 0<t<t_+\\
\frac{T_-}{2\sigma_{\rm S}}+\frac{\sqrt{\sigma_{+}}-\sqrt{\sigma_{-}}}{\mu t_-\sqrt{\sigma_{\rm S}}},~~ t_+<t<t_{\rm p}
\end{cases}
\label{expression-k-sise}
\end{equation}
with
\begin{equation}
\sigma_{\rm S}=
\begin{cases}
\frac{\sigma_{-}}{2}\left[1+(\sqrt{\frac{\sigma_{+}}{\sigma_{-}}}-1)
\frac{t}{t_+}\right]^2, 0<t<t_+\\
\frac{\sigma_{+}}{2}\left[1+(\sqrt{\frac{\sigma_{-}}{\sigma_{+}}}-1)
\frac{t-t_+}{t_-}\right]^2, t_+<t<t_{\rm p}
\end{cases}.
\label{eq:Seifert1}
\end{equation}
\textcolor{black}{
However, this protocol is no longer optimal when one imposes just maximum and minimum values on the response, i.e., $\sigma(t)\in[\sigma_-, \sigma_+]$. Then, our analysis shows that the maximum-efficiency and maximum-power protocol is still of the above form, but with $\sigma_-<\sigma(0)=\sigma(t_{\rm p})<\sigma(t_+)<\sigma_+$.}

In Fig.~\ref{SI-uni1}, we show the maximum power (a) and maximum efficiency (b) for the trial protocols under the constraint $\sigma_-\equiv\sigma(0)=\sigma(t_{\rm p})$ and $\sigma_+\equiv\sigma(t_+)$. As expected, power and efficiency corresponding to the protocol $\lambda_{\rm S}$ are largest from all the protocols. In particular, the figure demonstrates that the linear protocol, which was found to maximize output power for constrained $\lambda$, yields smaller output power than $\lambda_{\rm S}$. And the piecewise constant protocol yields smaller efficiency than $\lambda_{\rm S}$. Nevertheless, it is interesting to note that the performance of the protocol $\lambda_{\mathrm{slow}}(t)$, which optimizes both output power and efficiency for slow driving (see Sec.~\ref{sec:nonlinear} for details), is for the chosen parameters indistinguishable from that of $\lambda_{\rm S}$. This means that the chosen cycle is slow enough. Finally, for small enough cycles (small $\sigma_-/\sigma_+$) performance of all the protocols equal.

\begin{figure}[t]
\centering
\includegraphics[trim={0cm 0.8cm 2.3cm 0cm},width=0.95\columnwidth]{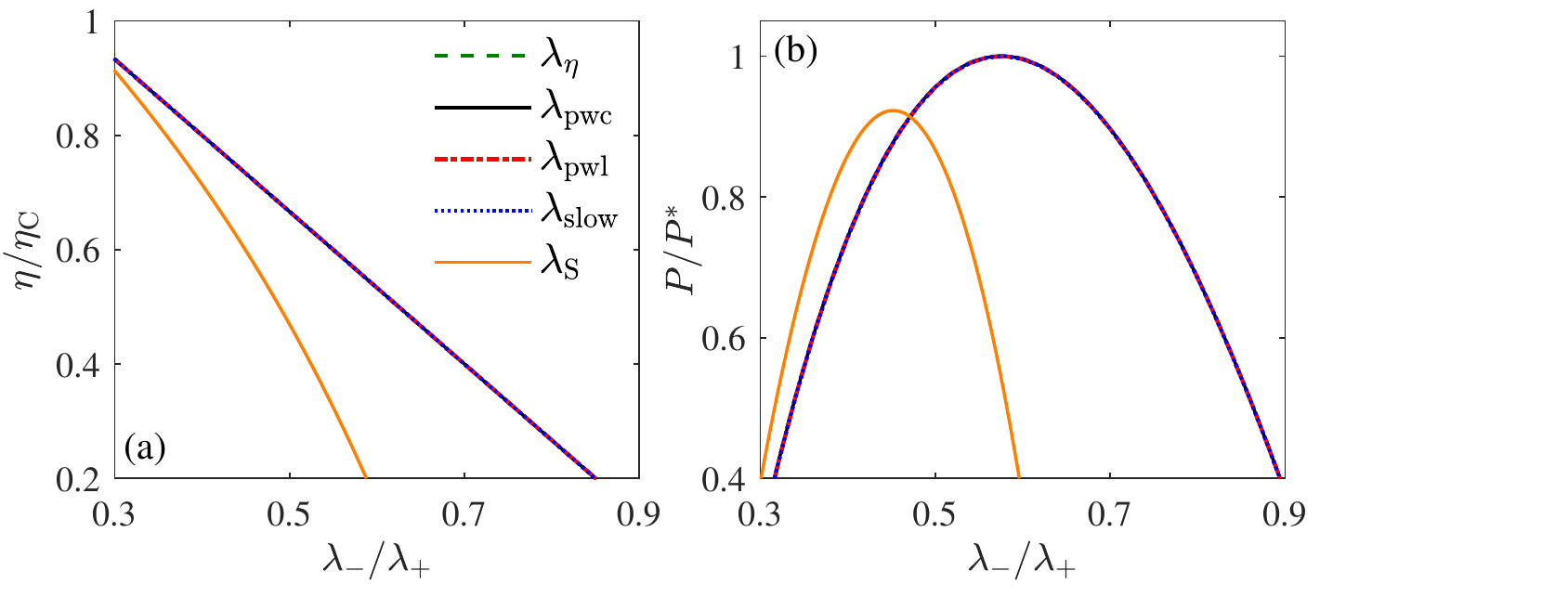}
\caption{(a) Maximum efficiency and (b) the corresponding power (in units of the ultimate maximum power $P^*$ \textcolor{black}{for $\lambda_{\rm pwl}$}) as functions of $\lambda_-/\lambda_+$. 
\textcolor{black}{
All protocols except for $\lambda_{\rm S}$ perfectly overlap. We used the same parameters as in Fig. 2 in the main text.}}
\label{SI-uni2-effi}
\end{figure}


\begin{figure*}[ht]
\centering
\includegraphics[trim={0cm 2.5cm 4cm 0cm},width=0.78\textwidth]{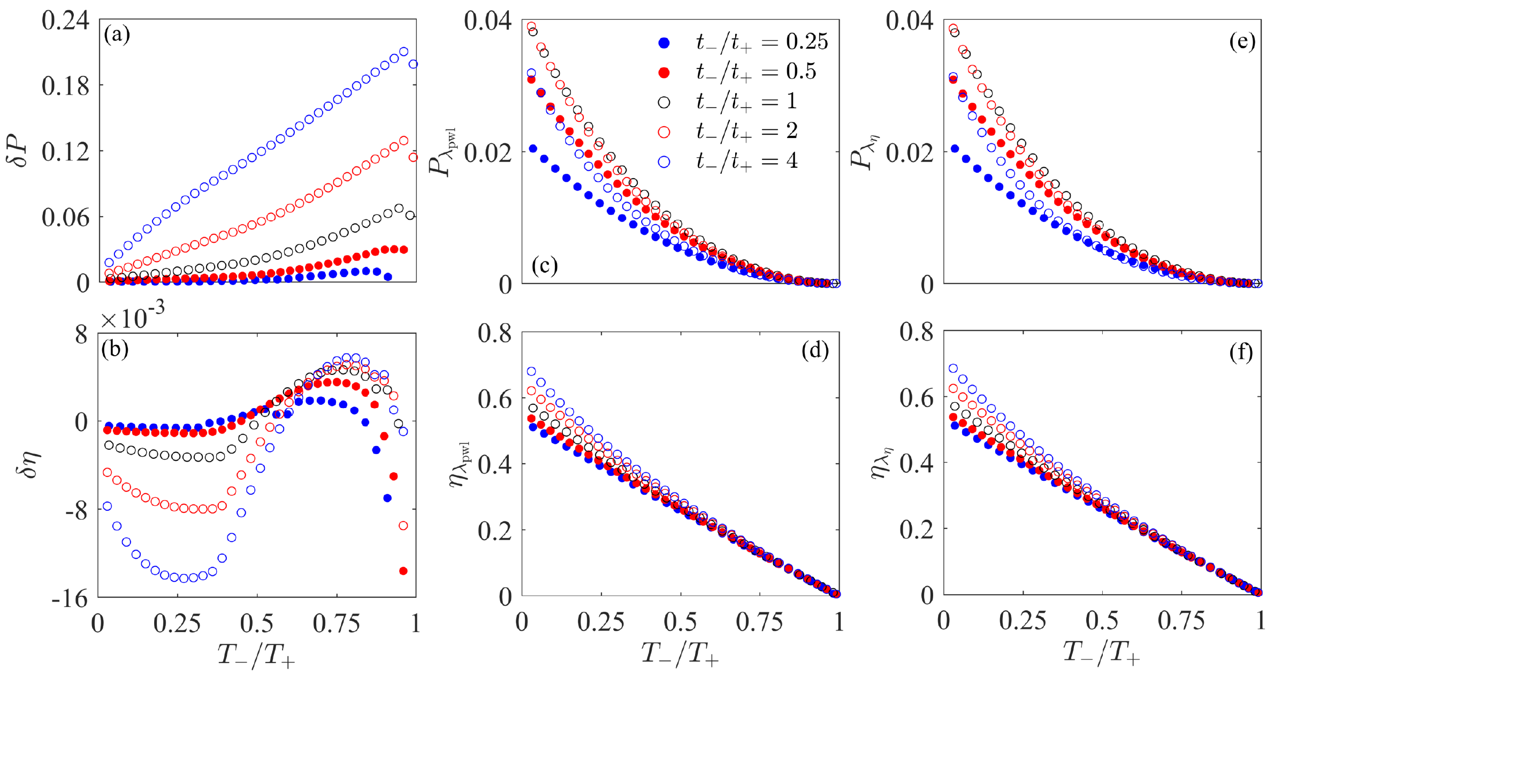}
\caption{
\textcolor{black}{
The relative differences $\delta X= (X_{\lambda_{\rm pwl}}-X_{\lambda_{\eta}})/X_{\lambda_{\rm pwl}}$ of (a) maximum power ($X=P$) optimized with respect to $\lambda_-$ and (b) the corresponding efficiency ($X=\eta$) for the linear protocol $\lambda_{\rm pwl}$ and the maximum-efficiency protocol \eqref{expression-sm-piece-protocol} for different values of $T_-/T_+$ and $t_-/t_+$. (c)-(f) show the corresponding values of maximum power and efficiency. The piecewise constant protocol $\lambda_{\rm pwc}$ and the maximum-efficiency protocol $\lambda_{\eta}$ \eqref{expression-sm-piece-protocol} are in this case equal. We used the same parameters as in Fig. 2 in the main text.}}
\label{fig:diff}
\end{figure*}

\subsection{Constrained control}
\label{SI-constrained-control-I}

Let us now turn to the optimisation problem with constraints on the control: $\lambda(t) \in [\lambda_{-}, \lambda_{+}]$. As we know from the main text, in this case the maximum-efficiency protocol is piecewise constant and maximum-power protocol is piecewise linear. The numerical proof of the latter is given in Fig.~2 in the main text.  In Fig.~\ref{SI-uni2-effi}, we show the numerical proof of the former: the optimal performance of all the trial protocols in the figure is the same as all of them collapse to the maximum-efficiency protocol (parameters $b$ and $d$ for all the protocols equal to 0).


\textcolor{black}{
In Fig. 2 in the main text, we show that maximum powers of the protocols $\lambda_{\eta}$ \eqref{expression-sm-piece-protocol} and $\lambda_{\rm pwl}$ differ just by 1\%. This suggests that the maximum-efficiency protocol \eqref{expression-sm-piece-protocol} can be used to estimate the overall maximum power optimized with respect to $\lambda_-$. In Fig. \ref{fig:diff}, we provide additional evidence supporting this conclusion. We show that for a broad range of values of $T_-/T_+$ and $t_-/t_+$, the relative difference in maximum power for the piecewice linear and (piecewise constant) maximum-efficiency protocol is small. Furthermore, from panels (c)-(f) we conclude that the optimal ratio $t_-/t_+$ is between 1 and 2, which is in agreement with the results of Sec. \ref{SI-section-short-isotherm} [see Eq. \eqref{SI-fast-alpha-eff} below]. Thus, for branch durations that optimize output power, the relative difference $\delta P$ in (a) is always below 12\%, and it decreases with the temperature ratio.
}

\section{Optimal driving for systems close to equilibrium}
\label{sec:nonlinear}

In this section, we consider optimisation of a slowly driven heat engine based on an overdamped Brownian particle trapped in the power-law potential $H=\lambda(t)x^n/n$ with $n = 2,4,\dots$. \textcolor{black}{We use the temperature protocol from Eq.~\eqref{expression-sm-piece-protocol} and impose fix values of the response $\sigma$ (or, in the slow driving limit equivalently also the control $\lambda$) at the ends of the two isotherms.} Dynamics of the particle position is described by the Langevin equation
\begin{equation}
\dot{x} = - \mu \lambda(t) x^{n-1} + \sqrt{2 D(t)} \xi(t),
\label{eq:Langevinxn}
\end{equation}
where $D(t)=\mu k_{\rm B}T(t)$ denotes the diffusion coefficient. From Eq.~\eqref{eq:Langevinxn} and its formal solution
\begin{equation}
    x(t) = -\mu\int dt\, \lambda(t) x^{n-1}(t) + \sqrt{2D(t)} \int dt\,\xi(t),
\end{equation}
we find that $\left<x(t)\xi(t)\right> = \sqrt{D/2}$ and thus
\begin{equation}
\frac{d}{dt}\left<x^2(t)\right> =
- 2 \mu \lambda(t) \left<x^n(t)\right> + 2D.
\label{eq:nonlinearmoments}
\end{equation}
Let us now assume that the control parameters $\{T(t),\lambda(t)\}$ vary slowly with respect to the relaxation time of the system, such that, during the limit cycle, the system is always close to equilibrium, and solve this equation up to the first order in $\dot{\lambda}(t)$. To this end, we consider the ansatz
$\left<x^2(t)\right> = \left<x^2(t)\right>_0$ and $\left<x^n(t)\right> = \left<x^n(t)\right>_0 + \left<x^n(t)\right>_{\dot{\lambda}}$, where
\begin{equation}
\left<x^m(t)\right>_0 = \int_{-\infty}^{\infty} dx x^m\, \frac{\exp(- \mu \frac{\lambda x^n}{n D})}{Z},
\end{equation}
with the partition function $Z = 2[n D/\mu \lambda(t)]^{1/n}\Gamma(1+1/n)$, is the value of the moment $\left<x^m(t)\right>$ corresponding to the infinitely slow driving, and $\left<x^n(t)\right>_{\dot{\lambda}}$ is the correction proportional to $\dot{\lambda}$. We find that
\begin{eqnarray}
\left<x^n(t)\right>_0 &=& \frac{D(t)}{\mu \lambda(t)},
\label{average-x-n-zero-fir}
\\    \left<x^2(t)\right>_0 &=& \left[\frac{n D(t)}{\mu \lambda(t)}\right]^{2/n} \frac{\Gamma(3/n)}{\Gamma(1/n)},
\end{eqnarray}
and
\begin{equation}
\left<x^n(t)\right>_{\dot{\lambda}} = - \frac{1}{2 \mu\lambda(t)} \frac{d}{dt} \left<x^2(t)\right>_0.
\end{equation}
We reiterate that this solution is valid only for protocols $\{T(t),\lambda(t)\}$ which are changing slowly with respect to the relaxation time of the system so that the system is during the whole cycle close to equilibrium. However, as we know from the previous discussion, both the piecewise constant maximum-efficiency protocol for constrained control and the optimal protocol~\eqref{expression-k-sise} for the constrained response contain discontinuities, where $\{T(t),\lambda(t)\}$ changes abruptly. To be able to use the slow driving approximation for the derivation of optimal cyclic protocols, we thus need to additionally assume that during these jumps, the system is not driven far from equilibrium. To this end, we assume that the ratio $\lambda(t)/T(t)$ in the Boltzmann factor is during the jumps at the ends of the isotherms constant. This additional assumption fixes the state of the system $\sigma$ at the ends of the isotherms and thus the present optimization scheme is only suitable for the optimization under the constrained response. Let us now proceed with the optimization.

Work done on the system during the time interval $t_{\rm i}\le t\le t_{\rm f}$ for the given Hamiltonian reads
\begin{equation}
W = \frac{1}{n}\int_{t_{\rm i}}^{t_{\rm f}} dt\, \dot{\lambda}(t) \left<x^n(t)\right> \equiv W(t_{\rm i},t_{\rm f}). 
\label{eq:worknonlinear}
\end{equation}
Having fixed the state of the system at the ends of the isotherms, it is enough to maximize the work during these branches. For an isothermal process, the work Eq.~\eqref{eq:worknonlinear} can be written as $W = \Delta F + W_{\rm irr}$, where  the first term, denoting the nonequilibrium free energy difference~\cite{Schmiedl2008}, comes from $\left<x^n(t)\right>_0$, and the second term reads
\begin{multline}
W_{\rm irr} = \frac{1}{n}\int_{t_{\rm i}}^{t_{\rm f}} dt\, \dot{\lambda}(t) \left<x^n(t)\right>_{\dot{\lambda}}=\frac{1}{n^2\mu}\left(\frac{n D}{\mu}\right)^{2/n}\\ \times\frac{\Gamma(3/n)}{\Gamma(1/n)}
\int_{t_{\rm i}}^{t_{\rm f}} dt\, \dot{\lambda}(t)^2 \lambda(t)^{-2 (1+n)/n}.
 \label{non-linear-wirr}    
\end{multline}
As $\Delta F$ is fixed by the imposed boundary conditions on the state of the system $\sigma$, to maximize the output work $-W$ means to minimize the irreversible work $W_{\rm irr}$ as a functional of $\lambda(t)$. This leads to the
Euler-Lagrange equation
\begin{equation}
    \ddot{\lambda}(t)\lambda(t) - \frac{1+n}{n} \dot{\lambda}(t)^2 = 0,
\end{equation}
which has the general solution
\begin{equation}
\lambda_{\rm slow}(t) =  \frac{a}{(1+bt)^n}.
\label{non-linear-sti}
\end{equation}
We thus come to an interesting conclusion that the optimal slow protocol for the constrained response scales with the same exponent as the potential. The values of $a$ and $b$ can be expressed in terms of the boundary conditions for $\lambda_{\rm slow}(t)$, i.e., $\lambda_\slow(t_{\rm i})\equiv\lambda_{\rm i}$ and $\lambda_\slow(t_{\rm f})\equiv\lambda_{\rm f}$. The optimal slow protocol \eqref{non-linear-sti} then reads
\be \label{eq:slow1}
    \lambda_\slow(t) =\frac{\lambda (t_{\rm i})}{\left[1+\left(\sqrt[n]{\frac{\lambda (t_{\rm i})}{\lambda (t_{\rm f})}}-1\right)\frac{t-t_{\rm i}}{t_{\rm f}-t_{\rm i}}\right]^n}.
\ee
And the corresponding irreversible work and input work are given by
\begin{eqnarray}
&W_{\rm irr}=\frac{\frac{\Gamma(3/n)}{\Gamma(1/n)} \left[\frac{n D}{\mu\lambda_{\rm i}}\right]^{2/n}\left(\sqrt[n]{\frac{\lambda_{\rm i}}{\lambda_{\rm f}}}-1\right)^2}{\mu (t_{\rm f}-t_{\rm i})},\\
&W = \frac{\frac{\Gamma(3/n)}{\Gamma(1/n)} \left[\frac{n D}{\mu\lambda_{\rm i}}\right]^{2/n}\left(\sqrt[n]{\frac{\lambda_{\rm i}}{\lambda_{\rm f}}}-1\right)^2}{\mu (t_{\rm f}-t_{\rm i})} -\frac{D}{n\mu}\ln\frac{\lambda_{\rm i}}{\lambda_{\rm f}}.   
\end{eqnarray}
These results are valid for the individual isothermal branches of the cycle. Importantly, the obtained optimized values of the irreversible work are correct up to the order $1/(t_{\rm f}-t_{\rm i})$, which is their exact dependence on the process duration~\cite{Schmiedl2008}. These results are thus exact even though they were obtained from the approximate optimal protocol. According to Refs.~\cite{Schmiedl2008,Holubec2016}, these irreversible works determine the optimal performance of the engine under the constrains on $\sigma$, i.e., they give the maximum output work $W_{\rm out} = - W(0,t_{+}) - W(t_{+},t_p)$ and efficiency $\eta = W_{\rm out}/[T_h \Delta S - W_{\rm irr}(0,t_{+})]$ ($\Delta S$ is the increase entropy of the system during the hot isotherm). Also this performance is thus from the approximate analysis based on the slow driving obtained exactly.



\end{document}